# Low-temperature formation of platinum silicides on polycrystalline silicon

Kirill V. Chizh,[a,*] Vladimir P. Dubkov,[a] Vyacheslav M. Senkov,[b] Igor V. Pirshin,[b] Larisa V. Arapkina,[a] Sergey A. Mironov,[a] Andrey S. Orekhov,[c] and Vladimir A. Yuryev [a,*]

[a] A.M. Prokhorov General Physics Institute of the Russian Academy of Sciences, 38 Vavilov Street, Moscow 119991, Russia

[b] P.N. Lebedev Physical Institute of the Russian Academy of Sciences, 53 Leninsky Avenue, Moscow 119991, Russia

[c] A.V. Shubnikov Institute of Crystallography of Federal Research Center "Crystallography and Photonics" of the Russian Academy of Sciences, 59 Leninsky Avenue, Moscow, 119333 Russia

*Corresponding authors.
E-mail addresses: chizh@kapella.gpi.ru (K.V. Chizh), vyuryev@kapella.gpi.ru (V.A. Yuryev).

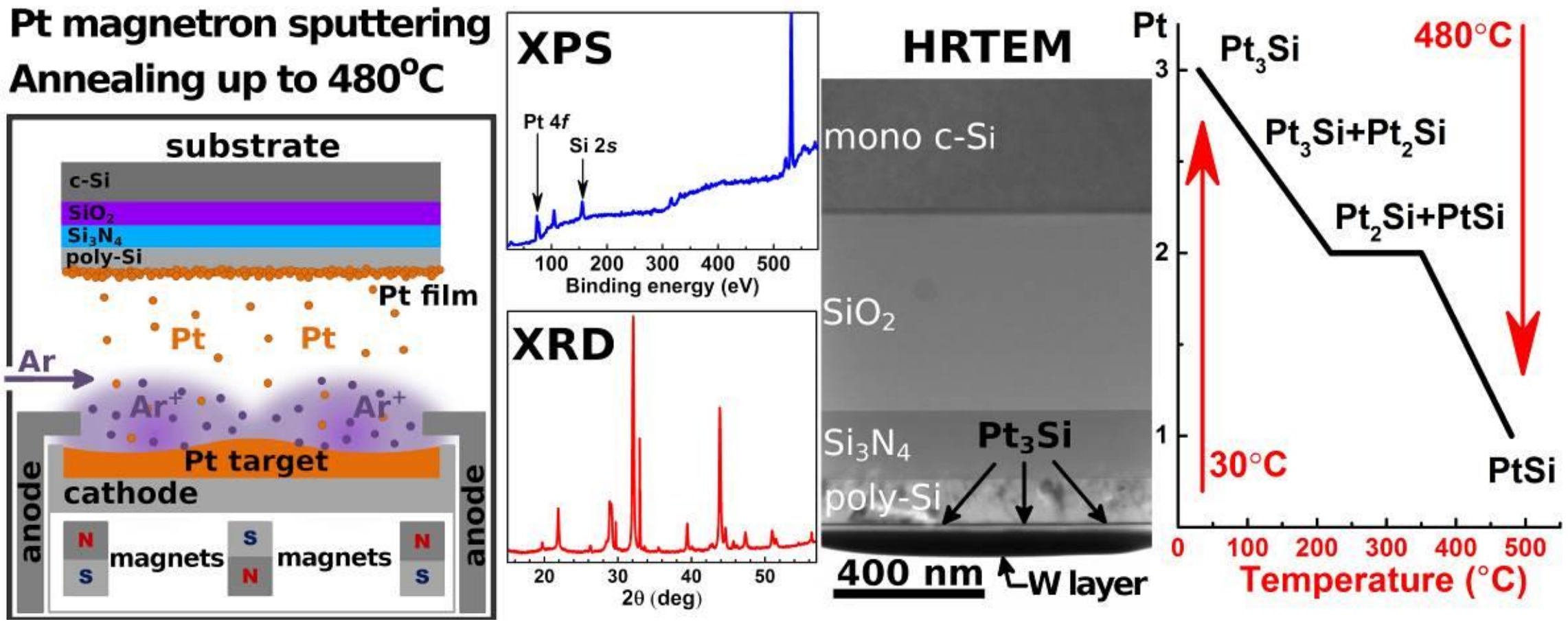

*Highlights*

- Interface domain in magnetron sputtered and annealed Pt/poly-Si film is studied
- $Pt_3Si$ and $Pt_2Si$ phases form at room temperature during Pt deposition on poly-Si
- Interfacial film relaxation turns $Pt_3Si$ to $Pt_2Si$ due to annealing at $100 < T < 200$°C
- $Pt_3Si$ and $Pt_2Si$ transit to PtSi due to annealing at temperatures from 320 to 480°C
- $Pt_3Si$ phase is detected up to the annealing temperature of 350°C

Abstract.

Phase formation and transitions in platinum silicide films on polycrystalline Si in the process of Pt magnetron sputtering and heat treatments at different temperatures are studied using X-ray photoelectron spectroscopy, high-resolution transmission electron microscopy and X-ray diffractometry. Phases of amorphous and polycrystalline $Pt_3Si$ and $Pt_2Si$ are shown to form during the room-temperature Pt deposition on poly-Si beneath the Pt layer. The relaxation of the interfacial film and partial transformation of $Pt_3Si$ into $Pt_2Si$ occurs as a result of the thermal treatment for 30 minutes in the temperature range from 125 to 300ºC. The $Pt_3Si$ and $Pt_2Si$ phases crystallize to PtSi due to annealing at the temperatures from 320 to 480ºC for the same time period.

## Introduction.

Platinum silicides compounds are widely applied in microelectronic industry due to their compatibility with the CMOS technology. The metallic nature of the conductivity, low electrical resistance and thermal stability make them an optimal material in the production of Schottky diodes [1,2], IR detectors [3–6], field-effect transistor gates, MEMS technology and the metallic conjunctions in microchips [1]. In the last decade, platinum silicides applications in the quantum wires [7], quantum dots [8] and nanostructures based on them were reported.

It is considered that the optimal platinum-silicide compound for industrial applications is PtSi that have the best thermal stability of the all silicides. Generally, the structures incorporating PtSi are obtained using thermal treatments of platinum layers deposited on silicon surfaces by magnetron sputtering [9] or electron gun evaporation [10]. However, it is known that, apart from platinum monosilicide, $Pt_2Si$ and $Pt_3Si$ phases also form under the annealing at the temperatures above 450$^o$C; besides that, partially unreacted platinum remains after this treatment [11]. Additionally, the layers obtained have rough boundaries and nonuniform thickness that impairs the quality of the films produced.

It has been shown, however, that stepped thermal treatment at the temperature range from 200 to 550$^o$C improves electrical and structural properties of the platinum silicide layers to be obtained [12]. Hence, a detailed study of the annealing temperature effect on the chemical composition of the films forming at the Pt/poly-Si interface is of significant scientific and technological interest for understanding the processes of formation of platinum silicides at low temperatures.

In this paper, the phase composition and the structure of an interfacial layer, originating between a magnetron sputtered platinum film and a poly-Si layer formed on a c-Si/$SiO_2$/$Si_3N_4$ dielectric substrate, which is widely used in MEMS technology [13], during Pt sputtering and after annealing at the temperatures $100 < T < 500^o$C for 30 min, are studied using X-ray photoelectron spectroscopy (XPS), high-resolution transmission electron microscopy (HR TEM) and X-ray diffractometry (XRD).

## 1. Experimental details

### *1.1. Sample preparation.*

Structures with the Pt silicide/poly-Si films were formed on commercial Czochralski-grown single-crystalline silicon wafers (ρ = 12 Ωcm, (100)-oriented, p-type) coated by a 527 nm thick layer of $SiO_2$ formed by thermal oxidation and a 174 nm thick layer of pyrolytic $Si_3N_4$ (the dielectric layers simulated a supporting membrane of a bolometer cell [14]). A film of polycrystalline Si with the thicknesses of 125 nm was deposited on the $Si_3N_4$ surface by thermal decomposition of monosilane at the substrate temperature of 620°C; then it was doped by implantation of phosphorus ions ($E_{P^+}$=60 keV) to the dose of $1.25\times10^{14}$ $cm^{-2}$ and annealing at 850°C for 30 min. A (20–22)-nm thick layer of platinum was deposited upon the poly-Si layer by magnetron sputtering at room temperature. Platinum silicide was formed by annealing in a quartz tube furnace at the different temperature from 125 to 480°C for 30 min in the atmosphere of Ar. For all samples, except for one (sample 11, Table 1), excess platinum was removed by chemical etching in a warm aqueous solution of *aqua regia* $H_2O$ : HCl : $HNO_3$ [4 : 3 : 1] for about 15 min. [6,13,15,16]

### *1.2. Experimental techniques and instruments*

#### *1.2.1. X-ray photoelectron spectroscopy*

The samples were studied by means of the X-ray photoelectron spectroscopy (XPS). The study was held in the analytical chamber of the Riber SSC2 UHV surface science center, equipped with cylindrical mirror type (CMA) electron energy analyzer (Riber EA 150). Residual gas pressure was $P \leq 5\times10^{-8}$ Pa. Photoexcitation was performed by means of the non-monochromatic Al $K_{\alpha1,2}$ X-ray radiation ($h\nu$ = 1486.6 eV) generated at the source power of 180 W.

All the photoelectron spectra were recorded at the fixed analyzer transmission (FAT) working mode with the pass energy $\Delta E_P = 16.4$ eV for the survey scans and $\Delta E_P = 5.6$ eV for the high-resolution spectra of specific elements. The instrumental resolution in these two cases was 1.8 eV and 0.62 eV, respectively.

The Pt 4*f* photoemission spectrum of each sample was deconvoluted into a few components related to different platinum silicides. The procedure of decomposition was held with XPSPEAK 4.1 program, which approximates photoelectron peaks by a product of Gaussian and Lorentz functions of the same width. Herewith each component was composed of the two similar parts, having an area ratio of 2 : 1 and separated by the energy $\Delta E = 0.5$ eV corresponding to the electron photoexcitation by the Al $K_{\alpha 1}$ and Al $K_{\alpha 2}$ components of the Al $K_{\alpha 1,2}$ radiation arising due to the spin-orbit splitting of the Al 2*p* orbital. The secondary electron background was calculated according to Shirley-Sherwood procedure [17,18] but in the case of a low signal-to-noise ratio linear background was used.

### *1.2.2. High-resolution transmission electron microscopy*

Transmission electron microscopy analysis was performed using FEI TITAN 80-300 microscope equipped with an Cs probe corrector and operated at the 300 kV accelerating voltage. Analysis of the samples was performed at conventional bright field (BF TEM) and dark field mode (DF TEM), as well as diffraction and scanning transmission mode (STEM). Energy dispersive X-ray spectroscopy (EDS) analysis was carried out using EDAX Si (Li) detector with energy resolution of 140 eV.

Thin TEM samples were prepared by focused ion beam (FIB) using FEI Helios NanoLab DualBeam scanning electron microscope (SEM). To protect the surface layer of the sample, a W-protection layer was deposited before the FIB procedure. Final polishing of the FIB samples was carried out by 2 kV $Ga^+$ ions using a low ion current.

### *1.2.3. X-ray diffractometry*

X-ray phase analysis was carried out with a D2 PHASER diffractometer (Bruker). An X-ray tube with a copper anode served as a radiation source. The signal was recorded by a GaAs linear detector array. The measurements were carried out in the ranges of diffraction angles $2\theta$ = 15 to 60º and 71 to 115º. At angles of $2\theta$ in the range 60 to 70, diffraction is significantly enhanced due to the presence of a Si (004) reflex related to a single-crystalline silicon substrate in this region. The linear semiconductor detector of the used diffractometer cannot work at such high X-ray intensities due to the generation of radiation defects in it. The scanning step along the angle $\Delta(2\theta)$ was 0.02º. The exposure at the point was 19 to 30 s. A 100 μm wide collimator slit was used. The recording was carried out without a Ni filter that reduces the signal intensity by several times. Reflexes were identified using the PDF-2 and ICSD databases [19–26].

## 2. Experimental Results

### *2.1. X-ray Photoelectron Spectroscopy*

The chemical composition and the structure of the interface, originated when platinum layers have been deposited on the poly-Si surface at the room temperature, and their changes under the thermal treatment up to 480°C was studied by means of the X-ray photoelectron spectroscopy. Figure 1 shows a typical survey XPS spectrum. High-resolution XPS spectra obtained close to the Pt 4*f* band from the samples processed at different temperatures are presented in Figure 2.

The survey photoelectron spectra for all the films, recorded at each film after chemical etching of the unreacted platinum, reveal peaks inherent to platinum, silicon, carbon and oxygen (Figure 1).

Detailed information on the platinum compounds contained in the interfacial layers was obtained by high resolution Pt 4*f* photoelectron line deconvolution. The Pt 4*f* peak of the each sample had been decomposed into a few components related to the different platinum silicides. Results obtained at different annealing temperatures are presented in Figure 2(a–i). Each component is a doublet with the peaks ratio 3 : 4, FWHM of 1.1 eV and the peak splitting energy of $\Delta E_b$ = 3.3 ± 0.05 eV.

It was found, that the doublet with the lowest electron binding energy ($E_b$ = 71.47 ± 0.04 eV) dominate in the Pt 4*f* line of the unannealed sample (Figure 2a). This value is close to the binding energies of the $Pt_3Si$ compound obtained in the works devoted to the synthesis of the platinum silicides. In Ref. [27], this energy is 71.45 eV, and the same value (71.45 ± 0.2 eV) is given in Ref. [28]. Thus, the component of the Pt 4*f* line discussed above may be attributed to the $Pt_3Si$ compound.

The next doublet shifted to the higher binding energy $E_b$ = 72.16 ± 0.02 eV is related to $Pt_2Si$ [27,28]. Since the first component makes the main contribution to the spectrum, the peak area ratio $S_{Pt3Si}/S_{Pt2Si}$ = 6.

The last component with the binding energy $E_b$ = 73.8 ± 0.3 eV can be associated with the $PtO_x$ ($x$ = 1, 2) oxide family [29]. Note that this component is present in the spectra of all samples; it emerges as a consequence of manipulations with the samples in ambient.

Low-temperature annealing of samples (125 ≤ $T$ ≤ 250°C) causes significant changes in the structure of the photoelectron peak (Figure 2b–d).

After annealing at the initial temperature of 125°C the contributions of the $Pt_3Si$ and $Pt_2Si$ silicides to the Pt 4*f* peak goes equal and peak area ratio $S_{Pt3Si}/S_{Pt2Si}$ = 1.07 ± 0.17 remains stable till the annealing temperature $T$ = 214°C (Figure 2d). With increasing annealing temperature, the contribution of the signal from $Pt_3Si$ to the spectrum begins to decrease, and at $T$ = 250°C the peak area ratio becomes $S_{Pt3Si}$/ $S_{Pt2Si}$ = 0.64 (Figure 2e).

Two new appreciable components appear in the Pt 4*f* line of the sample treated at $T$ = 300°C (Figure 2f). The first one, located at the binding energy $E_B$ = 72.60 ± 0.05 eV, is obviously related to platinum monosilicide, and a doublet attributed to the pure platinum has $E_B$ = 70.90 ± 0.02 эВ [27–29]. At the same time, the peak area ratio becomes $S_{Pt3Si}$/ $S_{Pt2Si}$ = 0.45, indicating further decrease of the $Pt_3Si$ content in the film.

The $Pt_2Si$ line dominates the photoelectronic spectrum obtained from the sample annealed at $T$ = 350ºC (Figure 2g), while the $Pt_3Si$ share decreases, and the PtSi contribution increases slightly as compared to the previous sample (Figure 2f).

High-temperature treatment $T \geq$ 400ºC (Figure 2h) leads to further changes in the structure of the Pt 4*f* spectrum. At $T$ = 400ºC, the component corresponding to the PtSi increases significantly and somewhat exceeds the contribution of the $Pt_2Si$ component. The $Pt_3Si$ signal drops to negligible values and there is no pure platinum signal. The Pt 4*f* line of the sample, annealed at $T$ = 480ºC (Figure 2i), indicates the last stage of the PtSi film formation on the bulk poly-Si border. The platinum monosilicide doublet dominate the spectrum. In addition, weak components attributed to $Pt_2Si$ silicide and the platinum oxide family are observed with approximately equal fractions ~ 0.1 (Figure 2i).

### *2.2. High-resolution transmission electron microscopy.*

The cross-section of the sample after magnetron sputtering and etching of Pt in the *aqua regia* solution (sample 1, Table 1) was examined by TEM methods. On the bright field TEM image, presented in Figure 3a, can be seen the microstructure of the sample with a well-ordered layered structure. It is seen that the $Pt_3Si$ layer located on top of poly-Si layer is continuous and uniformly distributed over a large field of view.

Analysis of the crystallinity of the $Pt_3Si$ layer was carried out by HRTEM technique. In Figure 3b the interface of poly-Si/$Pt_3Si$ layers is presented. The lattice planes of the poly-Si layer are clearly visible, while a disordered structure is observed for the $Pt_3Si$ layer. Local analysis of HRTEM image performed by the Fast Fourier Transform (FFT) method confirmed that the $Pt_3Si$ layer is amorphous.

### *2.3. X-ray Diffractometry*

The samples can be subdivided into three groups depending on features of their diffraction patterns: (I) the group 1 includes the samples subjected to heat-treatment at the temperature up to 300°C, including the unannealed one, followed by etching the excessive Pt off (samples 1 to 6 in Table 1), (II) the group 2 includes the samples subjected heat-treatment at the temperature from 320 to 480°C followed by etching the excessive Pt off (samples 7 to 10 in Table 1), and (III) the group 3 includes only the sample coated with Pt layer not subjected to further heat treatment or etching (sample 11 in Table 1).

Figure 4 demonstrates typical diffraction patterns obtained at the samples of the group 1 at the range of diffraction angels $2\theta = 20$ to 58º. Intense reflexes of Si are observed in the patterns. Besides, a weak peak is present around $2\theta = 39.8$º that may be attributed to reflexes of monoclinic $Pt_3Si$ and cubic Pt (Table 2). Similar diffraction patterns were obtained at the samples 3 and 5. A low-intensity peak related to PtSi is seen at $2\theta = 22.07$º in the sample 1, and a peak of polycrystalline $SiO_2$ is seen at $2\theta = 26.56$º in the sample 4.

A diffraction pattern of the sample 1, typical for all the samples of the group 1 in the range $2\theta = 70$ to 120º, is presented in Figure 5. Only reflexes related to polycrystalline silicon at the $CuK_\alpha$ и $CuK_\beta$ X-ray lines are seen in the pattern (Table 2).

The decaying signal in the range from 70 to 90º (Figure 5) is likely connected with the low-energy wing of a strong Si (004) band peaked at $2\theta = 69.197$º at the $CuK_\alpha$ line. The same reason gives rise to the increasing signal at $2\theta > 50$º (Figure 4) with approaching to the Si (004) $CuK_\beta$ reflex peaked at $2\theta = 61.69$º.

The tabulated positions of the reflexes from the PDF-2 and ICSD databases [19–26] and the data on the X-ray reflexes of $CuK_\alpha$ and $CuK_\beta$ lines obtained from the samples 1, 4 and 6 are presented in Table 2.

Some weak reflexes were always observed in the diffraction patterns; their signal-to-noise ratio ranged from ~1 to ~2. To check if they were due to an apparatus effect diffraction patterns of Si(001) wafers without platinum were recorded. These reflexes turned out to be present in these diffraction patterns and did not coincide with silicon ones. Therefore, we consider them as instrument related peaks and designate as ‘Instrument’ in the tables.

The presented data demonstrate that we have registered broad peaks, related to polycrystalline silicon of the structure, and narrow reflexes of $CuK_{\alpha1}$ and $CuK_{\alpha2}$ bands from (002) and (006) planes of Si connected with the substrate. The Si (002) reflex is forbidden; it is 3 orders of magnitude weaker than the allowed Si (004) reflex. Weak diffraction peaks around $2\theta = 39.8$ and 35.81° are observed for $CuK_\alpha$ and $CuK_\beta$ lines, respectively. These maximums may be connected with the presence of both Pt and $Pt_3Si$ since their peaks superimpose. Previously, for the identification of the cubic Pt phase, diffractograms of the Pt layer have been measured near the diffraction angle $2\theta = 85.80$° and an intense reflex of Pt (222) of cubic crystal system has been detected [13]. In this work, the peak at $2\theta = 85.80$° is not revealed for the group 1 samples (Figure 5) that evidences that the observed peaks at $2\theta = 39.80$ and 35.81° are related to $Pt_3Si$. These reflexes are more intense for the sample 1 but much weaker for the samples 2 to 6. Therefore, it may be concluded that some layer of crystalline $Pt_3Si$ remains in these samples — unannealed and annealed at $T \leq 300$ºC — after the platinum removal. Additionally, some amount of polycrystalline PtSi is contained in the sample 1; no polycrystalline $SiO_2$ is detected in the sample 6.

It should be noted that reflexes from $Si_3N_4$ and $SiO_2$ layers of the substrates are not observed in the diffraction patterns of the studied samples; evidently, they are amorphous. However, as mentioned above, a weak peak of polycrystalline $SiO_2$ is seen at $2\theta = 26.56$° in the sample 4. Likely, some quantity of crystalline $SiO_2$ has been synthesized in the sample, assumingly during Pt etching due to nitric acid contained in *aqua regia*.

Diffraction patterns of the group 2 samples annealed at the temperatures from 320 to 480ºC (samples 7 to 10, Table 1) strongly differ from the described above diffraction patterns of the group 1 samples (the diffraction patterns of the samples 7 and 10 at $2\theta = 17$ to 58° are plotted in Figure 6). Intense reflexes of platinum silicides, mainly of PtSi, emerge in these samples (Table 3). Reflexes related to $Pt_3Si$ and $Pt_2Si$ nearly are not observed in them that is likely connected with a fast reaction of the PtSi formation from $Pt_3Si$ and $Pt_2Si$ at these temperatures in a thin near-surface Pt/poly-Si interface layer. The results obtained for all the group 2 samples are qualitatively identical.

It is noteworthy that the structure of the studied layer considerably changes with the temperature increasing from 300 to 320ºC: strong reflexes of platinum silicides, mainly, of PtSi, appear in the diffraction patterns. A sharp change in the phase structure with temperature changes in a narrow range was also observed in the study of nickel and platinum silicides in Ref. [30] and was explained by the nucleation effect.

Figure 7 demonstrates the diffraction patterns of the samples 7 and 10 in the angle range $2\theta = 72$ to 115º. There is no peak at $2\theta = 85.80$º related to platinum in the curves that confirms the formation of PtSi in these samples.

The data on the reflexes of $CuK_\alpha$ and $CuK_\beta$ lines observed in these samples are gathered in Table 3. Reflexes peaking in the range $60º \leq 2\theta \leq 71º$ are not tabulated. In addition, several cells in Table 3 remain blanks due to the fact that the corresponding reflexes of the $CuK_\alpha$ or $CuK_\beta$ X-ray lines fall in the range of angles $60º \leq 2\theta \leq 71º$ where the strong line of the silicon substrate is located. Besides, note that the Si (006) reflex was not observed in the diffractogram of the sample 7 and is not given in the table.

To identify crystal phases forming in the process of Pt magnetron sputtering on poly-Si, diffraction patterns of a sample coated with 22 nm thick platinum film have been recorded. The examined sample was subjected neither to annealing nor to etching after platinum deposition (sample 11, Table 1). The diffraction pattern of this sample obtained in the angle ranges $20º \leq 2\theta \leq 60º$ and $82º \leq 2\theta \leq 120º$ is plotted in Figure 8. Intense bands of Pt (111) peaked at $2\theta = 39.60$º and 35.59º and Pt (222) peaked at $2\theta = 85.30$º and 75.47º for the $CuK_\alpha$ or $CuK_\beta$ lines, respectively, are observed in this sample. Peaks related to silicon are observed as well. No X-ray reflections connected with Pt silicides were revealed against the background of intense peaks of platinum in this sample.

## 3. Discussion

The data obtained by XPS, STEM and X-ray diffraction have allowed us to reveal specific details of formation of platinum silicide interface layer on the poly-Si surface depending on the annealing temperature from 480°C and below. On the STEM micrograph of the cross section of the initial structure (Figure 3), a uniform layer of 3 to 5 nm thick with the atomically sharp borders is shown.[1] The formed film prevents further migration of atoms to the bulk, which explains the presence of unreacted platinum on the surface. Chemical removing of pure metal Pt enabled the determination of chemical composition of the interfacial layers using the XPS. It was obtained, that the layer formed after Pt deposition without annealing consists mainly of $Pt_3Si$, and in lesser degree, of $Pt_2Si$ compounds. This is consistent with our previous data [16] as well as the results of the works on interaction of very thin platinum films with the surface of single-crystalline silicon [31,32]. Changes in the valence band of the platinum layer structure dependent on its thickness, studied by ultraviolet photoelectron spectroscopy at the liquid nitrogen temperature [31] and with synchrotron radiation at the room temperature [32], indicate the presence of the strong interaction between Si and Pt atoms, migrating beneath the surface of a Si single crystal. This leads to the formation of a metal-type layer when more than 20 monolayers are deposited. In addition, it can be assumed that the penetration of Pt atoms into silicon is facilitated by their kinetic energy acquired during magnetron sputtering. The data obtained by the X-ray diffraction in this sample have shown that a part of the $Pt_3Si$ silicide forms a crystalline phase.

The apparent contradiction between the XPS data on the presence of intense $Pt_3Si$ and $Pt_2Si$ signals in these samples and the results of an X-ray phase analysis demonstrating the low content of $Pt_3Si$ and the absence of $Pt_2Si$ can be explained as follows. Probably, these compounds are mainly in the amorphous state. When measured by the XPS method, they are well observed, yet they are practically invisible during the X-ray phase analysis. This is indirectly confirmed by the TEM image (Figure 3) since rows of atoms are observed in region poly-Si, whereas rows of atoms are not visible in region $Pt_3Si$, therefore, the film is likely predominantly amorphous.

---

[1] Note, that this layer is chemically resistant to etching in *aqua regia*.

The absence of peaks assigned to $Pt_2Si$ in the X-ray diffraction patterns of unannealed sample with deposited and unetched Pt (sample 11, Table 1), which have been previously observed by us at the samples similar to the sample 11 and interpreted using X-ray reflection as a $Pt_2Si$ layer beneath the Pt layer [13], is due to insufficient sensitivity of the instrument used in this work in comparison with the two-band device employed in our previous research. The presence of $Pt_2Si$ in the sample 1, shown using XPS in this work, agrees with our previous data and confirms the presence of this compound under Pt after magnetron sputtering reported in [13].

It should be emphasized that the metal-rich interfacial layer formed at room temperature should not be stable for the two reasons. The first one is that platinum atoms included into the layer are in silicide-like coordination limited to the first nearest neighbors. This is indicated by a sharp silicide-silicon interface. Therefore, chemical binding in the layer obtained is weakened [33]. The next one is originated by significant volumetric shrinkages (reaching 0.15) of different platinum silicides compared to crystalline silicon. Hence, compressive stresses appear in the interfacial layer during platinum silicides formation on the poly-Si surface. The intrinsic compressive stresses may relax partly through the same diffusion processes that cause the silicides formation [34]. The diffusion processes may be significantly intensified due to the decrease of atomic bond breaking barrier, caused by atomic bond loading in the stressed domain, and their rupturing by local energy fluctuations in the analogy with the process of stressed solids fracture comprising, according to Ref. [35], a sequence of such elementary events.

Heat treatment of the initial samples at $T$ = 125ºC actually leads to relaxation of the compressive strain in the interfacial film discussed above. Indeed, equalization of the $Pt_3Si$ and $Pt_2Si$ signals in the Pt 4*f* photoemission peak indicates an increase in $Pt_2Si$ content in the interfacial layer, while $Pt_3Si$ phase content goes down. This is approved with X-ray diffraction data, under which $Pt_3Si$ signal ($2\theta$ = 39,80º) decreases. The absence of the $Pt_2Si$ phase peaks in the X-ray diffraction patterns is, as mentioned above, explained by the relatively low instrument sensitivity.

Phase composition of the interfacial film remains stable until the annealing temperature exceeds 214ºC. Noteworthy, that approximate equality of the $Pt_3Si$ and $Pt_2Si$ signals have been previously observed in the interfacial layer formed of platinum, deposited on the surface of monocrystalline silicon, after annealing at $T$ = 150ºC [36]. An increase of the annealing temperature to $T$ = 250ºC causes a decrease in the $Pt_3Si$ signal, which can be explained by the gradual transformation of this phase to the $Pt_2Si$ compound and by the diffusion of Pt atoms into the silicon bulk activated at this temperature.

In the sample annealed at $T$ = 300ºC, the $Pt_3Si$ content continues to decrease. In addition, the signal of pure platinum is observed in the Pt 4*f* spectrum. The appearance of unreacted Pt in the film probably results from heterogeneous nucleation of the phases near the silicide–silicon interface, similar to the process that occurred in the Pt/c-Si structure heat-treated at 240ºC [12].

The XPS spectrum of the Pt 4*f* core line of the sample 8 annealed at $T$ = 350ºC reveals the presence of the PtSi phase, while $Pt_3Si$ content sufficiently decreases. The signal of unreacted platinum is also reduced. A peak area ratio $S_{Pt3Si}/ S_{Pt2Si}$ is the same as that of the layer prepared as described in Ref. [36] and annealed at $T$ = 230ºC. Heat treatment of samples at $T$ = 400ºC causes a sharp increase in the PtSi signal, which begins to dominate the spectrum. Finally, Pt 4*f* spectrum of the sample annealed at $T$ = 480ºC almost entirely consists of the PtSi signal, which indicates the completion of the Pt/poly-Si interface. These data are in good agreement with the results of the X-ray diffractometry, according to which the diffraction patterns obtained from the samples of the group 2 heat-treated at the temperatures from 320 to 480ºC present intense peaks of PtSi, while the signals from $Pt_3Si$ and $Pt_2Si$ are not detected.

This study shows that the kinetics of PtSi film formation at the platinum–silicon interface under appropriate thermal treatment are very similar, if polycrystalline or monocrystalline silicon substrates are used, but there are significant differences as well. The first one is the presence of unreacted platinum on each poly-Si sample after annealing in this study, while on monocrystalline silicon, all platinum deposited under ultra-high vacuum conditions is consumed during heat treatment.[2] The next distinction is that the diffusion is activated at the temperature about 250ºC on the poly-Si surface, while this takes place at 190ºC to 200ºC for the monocrystalline Si substrates. Finally, the diffusion in the layers studied in this work is slower, compared to that occurring in the films formed on the monocrystalline Si substrates. This seems to be due to the presence of the multiple defects on the poly-Si surface. This effect may be driven by small concentration of oxygen atoms originated in the interface layer from grain boundaries and after manipulations in the ambient [11,37].

## 4. Conclusion

In summary, it is shown that the metal-enriched interfacial film formed on the surface of poly-Si beneath a layer of the magnetron sputtered platinum consists of $Pt_3Si$ and $Pt_2Si$ silicides, being in an amorphous and crystalline state.

Sample heating in the temperature range from 125 to 214ºC for 30 minutes causes the interfacial film relaxation via the partial transition of $Pt_3Si$ to $Pt_2Si$. Further increase in the annealing temperature leads to the further reduction of $Pt_3Si$ phase content in the layer; this phase is detected up to annealing temperature of 350ºC. At annealing temperatures from 320 to 480ºC, the transition of $Pt_3Si$ and $Pt_2Si$ phases into the PtSi phase occurs.

We explain the observed features of the formation of interfacial films between platinum and the poly-Si surface at different temperatures, characteristic to ploy-Si as distinct to single crystalline Si, by the existence of multiple defects in the polycrystalline Si substrate and the presence of some amount of oxygen atoms appeared in the layers at Pt/Si interface from grain boundaries and due to manipulations with the samples in the air ambient.

[2] It was this excess Pt coating that needed to be chemically removed in order to conduct measurements.

## Conflict of interest

None declared.

## Acknowledgements

We thank Mr. Oleg Y. Nalivaiko of JSC "Integral" (Minsk, Belarus) for the production of the artificial dielectric c-Si/$SiO_2$/$Si_3N_4$ substrates and the deposition of poly-Si and Pt and Mrs. Lyudmila A. Krylova of GPI RAS for chemical treatments of the samples. The authors would like to thank Alexey M. Lomonosov for proofreading and many useful insights. We also appreciate Mrs. Natalia V. Kiryanova of GPI RAS for her contribution to the project handling.

The Russian Foundation for Basic Research funded this work through the grant number 18-52-00033. We appreciate Mrs. Natalia V. Kiryanova of GPI RAS for her contribution to the project handling.

The Center for Collective Use of Scientific Equipment of GPI RAS supported this research via presenting admittance to its equipment.

TEM measurements were performed with financial support of the Ministry of Science and Higher Education of the Russian Federation within the state assignment for the Federal Scientific Research Centre "Crystallography and Photonics" of the Russian Academy of Sciences.

**References.**

[1] S.P. Murarka, Silicides For VLSI Applications, Academic Press, New York,1983.

[2] S.P. Murarka, Silicide thin films and their applications in microelectronics, Intermetallics 3 (1995) 173–186. https://doi.org/10.1016/0966-9795(95)98929-3.

[3] M. Kimata, Metal Silicide Schottky Infrared Detector Arrays, in: Capper P., Elliott C.T. (eds) Infrared Detectors and Emitters: Materials and Devices. Electronic Materials Series, vol. 8. Springer, Boston, MA, 2001: pp. 77–98. https://doi.org/10.1007/978-1-4615-1607-1_4.

[4] M. Hosseinifar, V. Ahmadi, M. Ebnali-Heidari, Si-schottky photodetector based on metal stripe in slot-waveguide microring resonator, IEEE Photonics Technol. Lett. 28 (2016) 1363–1366. https://doi.org/10.1109/LPT.2016.2543602.

[5] A.V. Voitsekhovskii, A.P. Kokhanenko, S.N. Nesmelov, S.I. Lyapunov, N. V. Komarov, Photoelectric properties of photodetectors based on silicon-platinum silicide schottky barriers with a highly-doped surface layer, Russ. Phys. J. 44 (2001) 1139–1151. https://doi.org/10.1023/A:1015393305423.

[6] V.A. Yuryev, K.V. Chizh, V.A. Chapnin, V.P. Kalinushkin, Metal silicide/Si thin-film Schottky-diode bolometers, Proc. SPIE 9519 (2015) 95190K. https://doi.org/10.1117/12.2178487.

[7] Y.C. Lin, K.C. Lu, W.W. Wu, J. Bai, L.J. Chen, K.N. Tu, Y. Huang, Single crystalline PtSi nanowires, PtSi/Si/PtSi nanowire heterostructures, and nanodevices, Nano Lett. 8 (2008) 913–918. https://doi.org/10.1021/nl073279r.

[8] S. Miyazaki, M. Ikeda, K. Makihara, K. Shimanoe, R. Matsumoto, Formation of metal silicide nanodots on ultrathin $SiO_2$ for floating gate application, Solid State Phenom. 154 (2009) 95–100. https://doi.org/10.4028/www.scientific.net/SSP.154.95.

[9] S.P. Murarka, E. Kinsbron, D.B. Fraser, J.M. Andrews, E.J. Lloyd, High

temperature stability of PtSi formed by reaction of metal with silicon or by cosputtering, J. Appl. Phys. 54 (1983) 6943–6951. https://doi.org/10.1063/1.332010.

[10] P.I. Gaiduk, A. Nylandsted Larsen, Platinum-silicide formation during rapid thermal annealing: Dependence on substrate orientation and pre-implanted impurities, Appl. Phys. A 53 (1991) 168–171. https://doi.org/10.1007/BF00323878.

[11] F. Nava, S. Valeri, G. Majni, A. Cembali, G. Pignatel, G. Queirolo, The oxygen effect in the growth kinetics of platinum silicides, J. Appl. Phys. 52 (1981) 6641–6646. https://doi.org/10.1063/1.328655.

[12] F. Komarov, O. Milchanin, T. Kovalyova, J. Solovjov, A. Turtsevich, C. Karwat, Low temperature formation of platinum silicide for Shottky diodes contact layer, in: 9th Int. Conf. "Interaction of Radiation with Solids," Minsk, Belarus, 2011, p. 367.

[13] V.A. Yuryev, K. V. Chizh, V.A. Chapnin, S.A. Mironov, V.P. Dubkov, O. V. Uvarov, V.P. Kalinushkin, V.M. Senkov, O.Y. Nalivaiko, A.G. Novikau, P.I. Gaiduk, Pt silicide/poly-Si Schottky diodes as temperature sensors for bolometers, J. Appl. Phys. 117 (2015) 204502. https://doi.org/10.1063/1.4921595.

[14] V.A. Yuryev, V.A. Chapnin, K. V Chizh, V.Y. Resnik, V.P. Korol'kov, G.A. Rudakov, Schottky barrier thermal diodes for uncooled microbolometric detectors of radiation, in: XXII International Scientific and Engineering Conference on Photoelectronics and Night Vision Devices, May 22–25, 2012, (Moscow, Russia, Orion Res. Prod. Assoc.) 2012, pp. 322–324.

[15] S.A. Mironov, V.P. Dubkov, K. V. Chizh, V.A. Yuryev, Room-temperature formation of $Pt_3Si/Pt_2Si$ films on poly-Si substrates, J. Phys. Conf. Ser. 816 (2017) 012011. https://doi.org/10.1088/1742-6596/816/1/012011.

[16] M.S. Storozhevykh, V.P. Dubkov, L.V. Arapkina, K.V. Chizh, S.A. Mironov, V.A. Chapnin, V.A. Yuryev, Silicon-germanium and platinum silicide nanostructures for silicon based photonics, Proc. SPIE 10248 (2017) 102480O.

https://doi.org/10.1117/12.2265882.

[17] D.A. Shirley, High-resolution x-ray photoemission spectrum of the valence bands of gold, Phys. Rev. B. 5 (1972) 4709–4714. https://doi.org/10.1103/PhysRevB.5.4709.

[18] D. Briggs, M.P. Seah (editors.), Practical Surface Analysis, Auger and X-ray Photoelectron Spectroscopy, Wiley, 1996. https://books.google.com/books?id=JClJAQAACAAJ.

[19] H. Lipson, H. Steeple, Interpretation of X-ray Powder Diffraction Patterns, Macmillan, New York, 1970.

[20] International Centre for Diffraction Data, Powder Diffraction File, Card No. 01-071-0523 (orthorombic PtSi).

[21] International Centre for Diffraction Data, Powder Diffraction File, Card No. 04-0802 (cubic Pt).

[22] International Centre for Diffraction Data, Powder Diffraction File, Card No. 86-1630 (hexagonal $SiO_2$).

[23] International Centre for Diffraction Data, Powder Diffraction File, Card No. 83-0152 (orthorombic PtSi).

[24] International Centre for Diffraction Data, Powder Diffraction File, Card No. 01-077-4407 (monoclinic $Pt_3Si$).

[25] International Centre for Diffraction Data, Powder Diffraction File, Card No. 17-0670 (monoclinic $Pt_3Si$).

[26] International Centre for Diffraction Data, Powder Diffraction File, Card No. 01-070-2057 (cubic Pt).

[27] F. Streller, G.E. Wabiszewski, F. Mangolini, G. Feng, R.W. Carpick, Tunable, source-controlled formation of platinum silicides and nanogaps from thin precursor films, Adv. Mater. Interfaces 1 (2014) 1300120.

https://doi.org/10.1002/admi.201300120.

[28] R.T. Fryer, R.J. Lad, Synthesis and thermal stability of $Pt_3Si$, $Pt_2Si$, and PtSi films grown by e-beam co-evaporation, J. Alloys Compd. 682 (2016) 216–224. https://doi.org/10.1016/j.jallcom.2016.04.260.

[29] A. V. Naumkin, A. Kraut-Vass, S.W. Gaarenstroom, C.J. Powell, NIST X-ray Photoelectron Spectroscopy Database, NIST Standard Reference Database Number 20, National Institute of Standards and Technology, Gaithersburg MD, 20899 (2012). https://doi.org/10.18434/T4T88K.

[30] P. Gas, F.M. D'Heurle, F.K. LeGoues, S.J. La Placa, Formation of intermediate phases, $Ni_3Si_2$ and $Pt_6Si_5$: Nucleation, identification, and resistivity, J. Appl. Phys. 59 (1986) 3458–3466. https://doi.org/10.1063/1.336815.

[31] I. Abbati, L. Braicovich, B. De Michelis, O. Bisi, R. Rovetta, Electronic structure of compounds at platinum-silicon (111) interface, Solid State Commun. 37 (1981) 119–122. https://doi.org/10.1016/0038-1098(81)90725-0.

[32] G. Rossi, I. Abbati, L. Braicovich, I. Lindau, W.E. Spicer, Si(111)-Pt interface at room temperature: A synchrotron radiation photoemission study, Phys. Rev. B. 25 (1982) 3627–3636. https://doi.org/10.1103/PhysRevB.25.3627.

[33] G. Rossi, d and f metal interface formation on silicon, Surf. Sci. Rep. 7 (1987) 1–101. https://doi.org/10.1016/0167-5729(87)90005-7.

[34] F.M. d'Heurle, P. Gas, Kinetics of formation of silicides: A review, J. Mater. Res. 1 (1986) 205–221. https://doi.org/10.1557/JMR.1986.0205.

[35] A.I. Slutsker, Atomic-level fluctuation mechanism of the fracture of solids (computer simulation studies), Phys. Solid State. 47 (2005) 801–811. https://doi.org/10.1134/1.1924836.

[36] J. Čechal, T. Šikola, A study of the formation and oxidation of PtSi by SR-PES, Surf. Sci. 600 (2006) 4717–4722. https://doi.org/10.1016/j.susc.2006.07.041.

[37] C. Chang, A. Segmüller, H.-C. W. Huang, B. Cunningham, F.E. Turene, A.

Sugerman, P.A. Totta, PtSi contact metallurgy using sputtered Pt and different annealing processes, J. Electrochem. Soc. 133 (1986) 1256–1260. https://doi.org/10.1149/1.2108830.

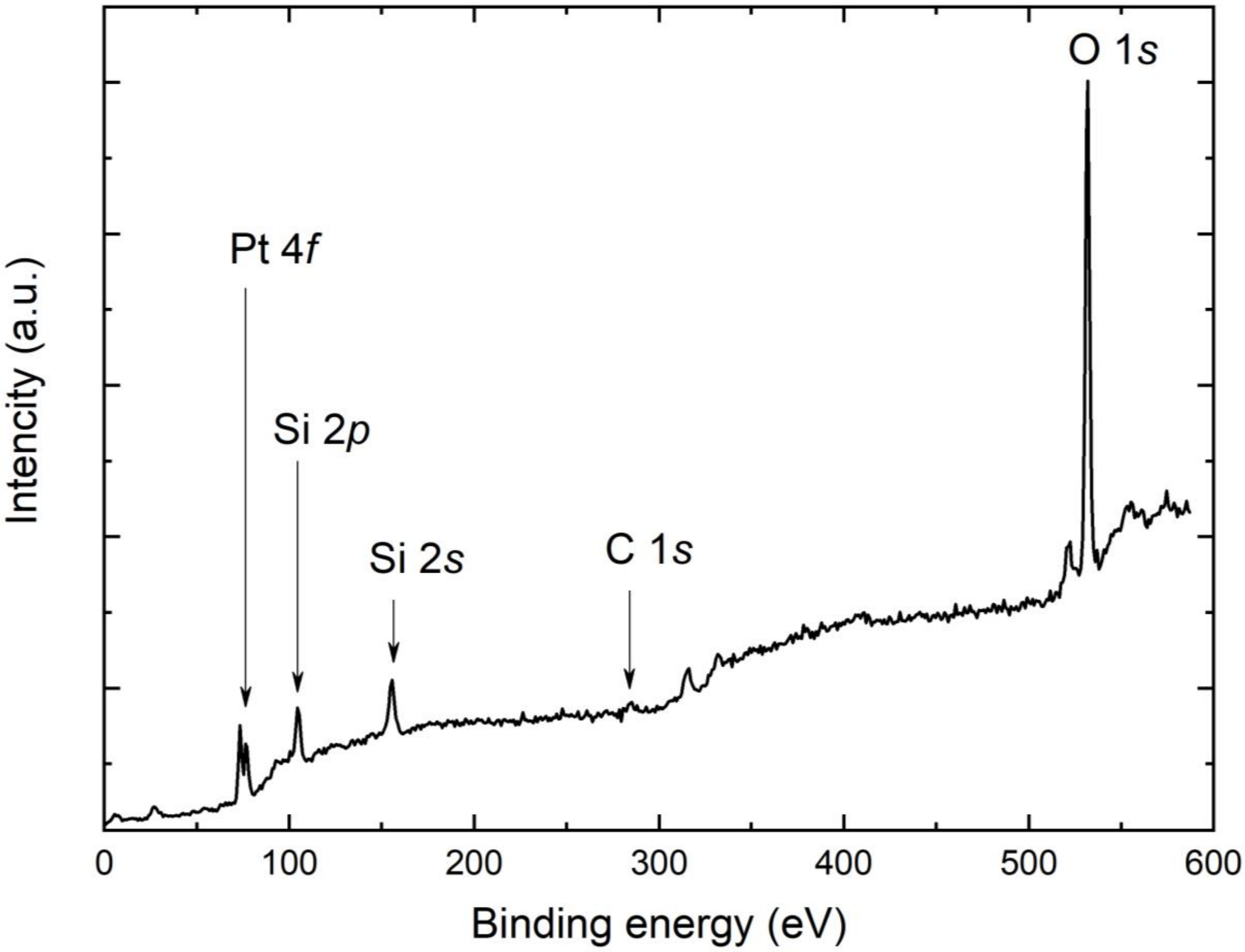

Figure 1. A survey spectrum of the sample 1 (Table 1).

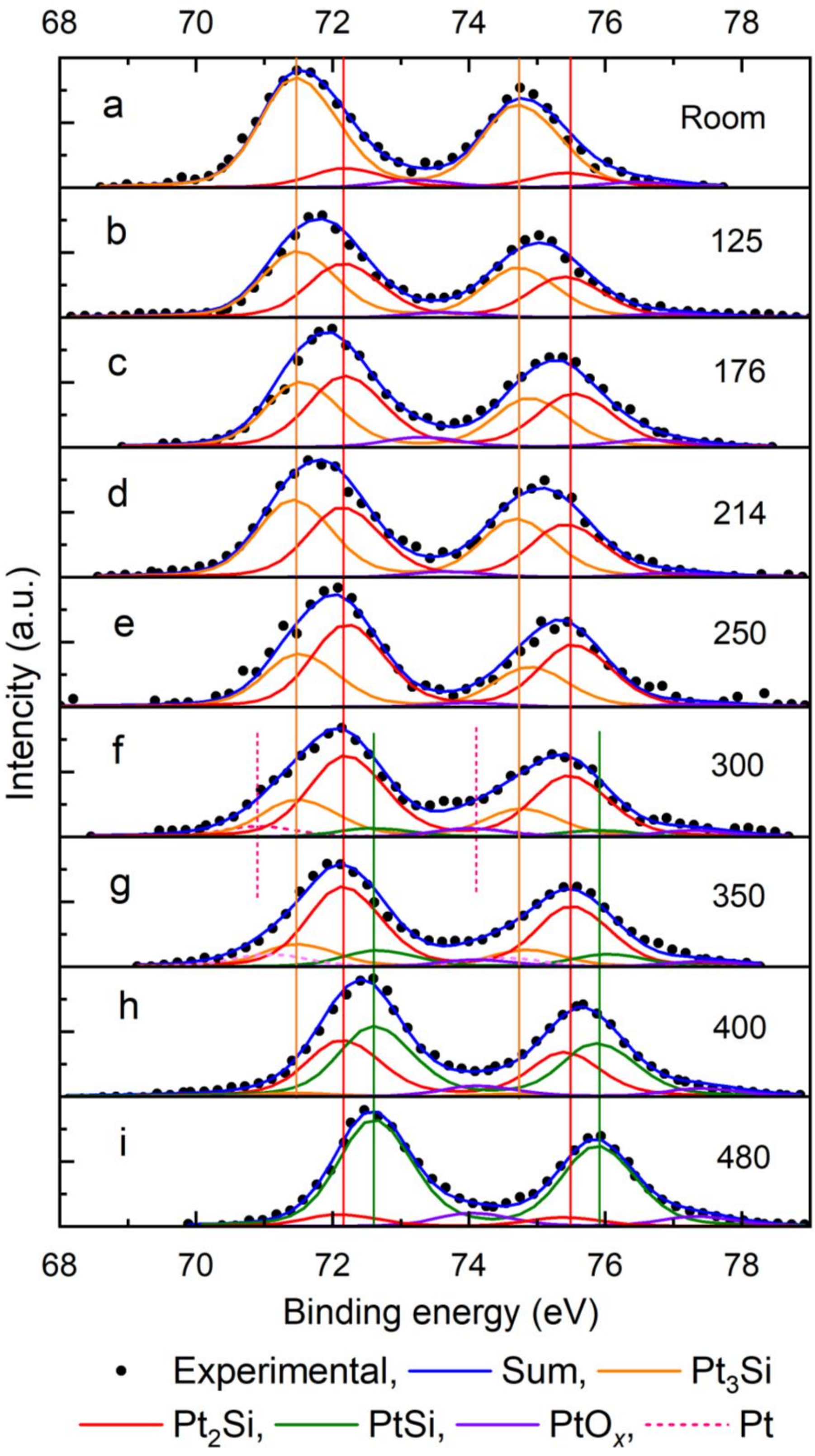

Figure 2. High-resolution XPS spectra of the Pt 4*f* lines from the samples thermally treated at the different temperatures (Table 1); the annealing temperatures are shown in the panels.

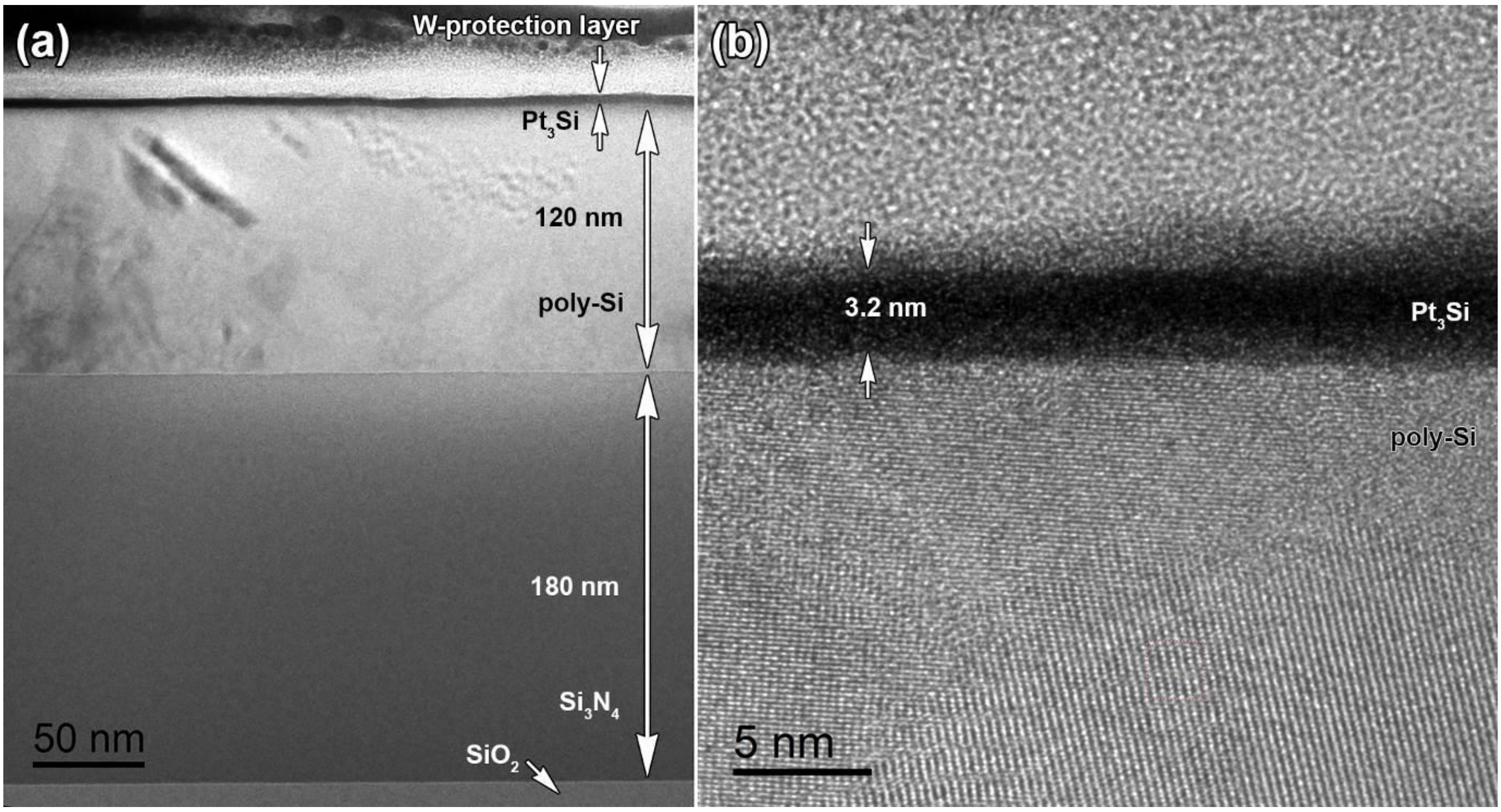

Figure 3. BFTEM image of a c-Si/$SiO_2$/$Si_3N_4$/poly-Si/ $Pt_3Si$ sample cross-section after magnetron sputtering of Pt followed by Pt liquid etching (a), HR TEM image of the $Pt_3Si$/poly-Si interface (b).

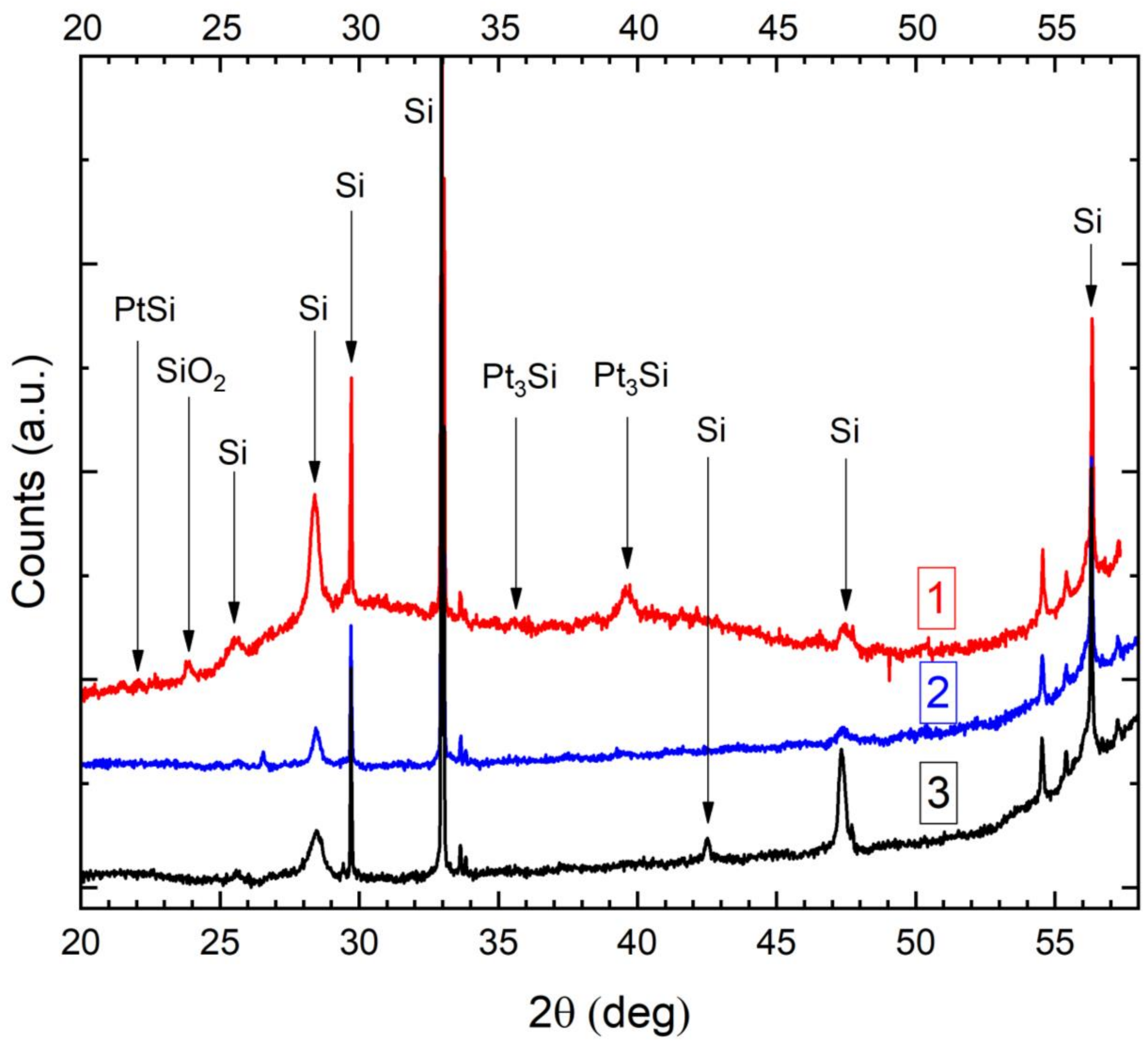

Figure 4. Diffraction patterns (mixed $CuK_\alpha$ and $CuK_\beta$) of the samples 1, 4 and 6 (Table 1) recorded within the angle range $2\theta$ = 17 to 58º: the unannealed sample (curve 1), and the samples heat-treated at 214°C (curve 2) and 300°C (curve 3); for clarity, the curves are offset vertically.

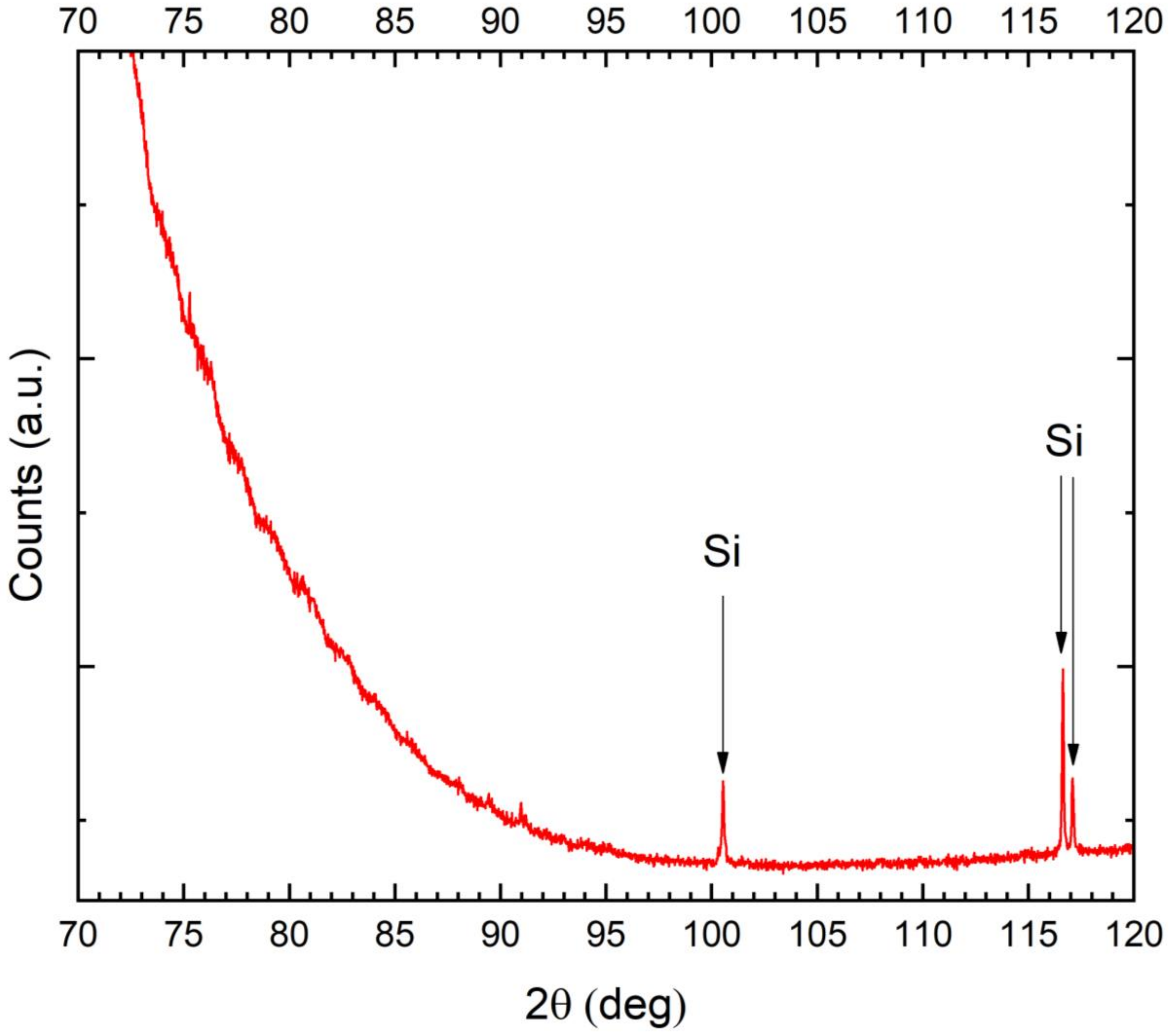

Figure 5. A diffraction pattern (mixed $CuK_\alpha$ and $CuK_\beta$) of the unannealed sample after Pt removal (sample 1, Table 1) within the high-angle range ($2\theta$ = 70 to120º).

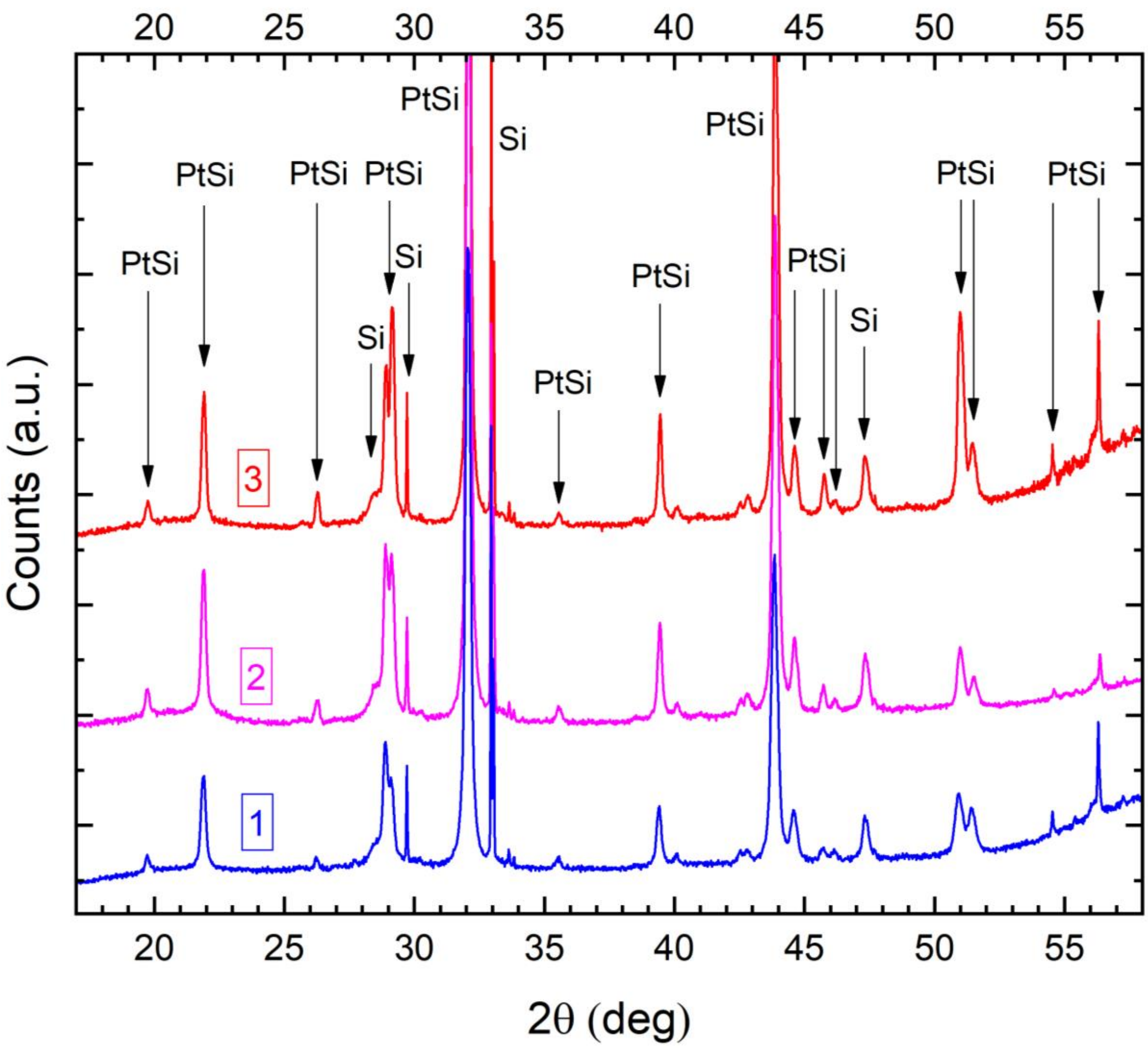

Figure 6. Diffraction patterns (mixed $CuK_\alpha$ and $CuK_\beta$) of the samples 7, 9 and 10 (Table 1) heat-treated at 320 (curve 1), 400 (curve 2) and 480°C (curve 3), respectively, recorded within the angle range $2\theta = 17$ to 58º; for clarity, the curves are offset vertically.

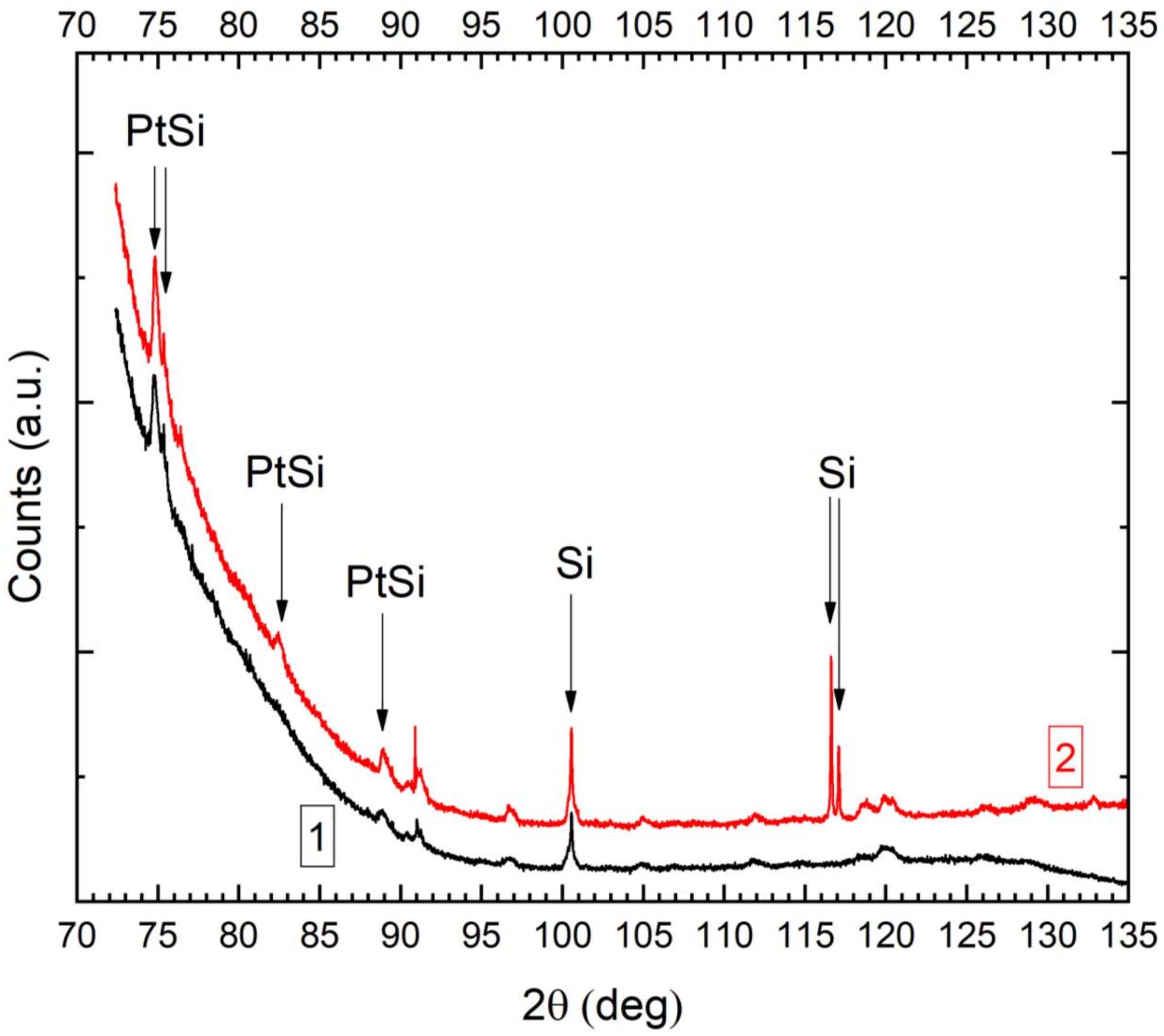

Figure 7. Diffraction patterns (mixed $CuK_{\alpha}$ and $CuK_{\beta}$) of the samples 7 and 10 (Table 1) heat-treated at 320 (curve 1), and 480°C (curve 2), respectively, recorded within the angle range $2\theta$ = 72 to 136º; for clarity, the curves are offset vertically.

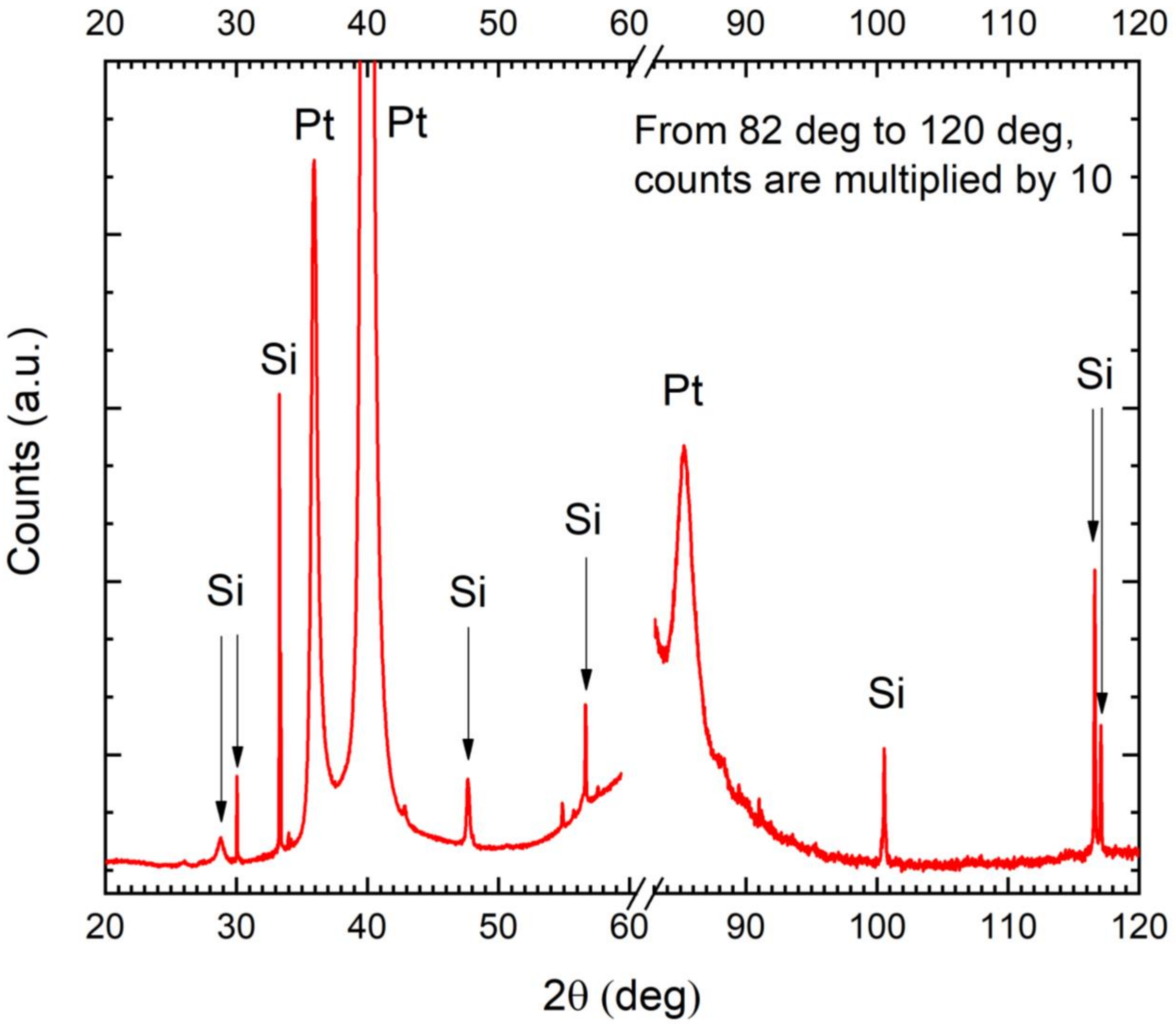

Figure 8. Diffraction patterns (mixed $CuK_{\alpha}$ and $CuK_{\beta}$) of the structure with the 22-nm thick Pt layer (sample 11, Table 1).

Table 1. Modes of heat treatments and etching of samples.

| Sample Number | Annealing temperature (ºC) | Annealing time (min) | Pt etching |
|---|---|---|---|
| 1 | As fabricated | - | yes |
| 2 | 125 | 30 | yes |
| 3 | 176 | 30 | yes |
| 4 | 214 | 30 | yes |
| 5 | 250 | 30 | yes |
| 6 | 300 | 30 | yes |
| 7 | 320 | 30 | yes |
| 8 | 350 | 30 | yes |
| 9 | 400 | 30 | yes |
| 10 | 480 | 30 | yes |
| 11 | As fabricated | - | no |

Table 2. XRD reflexes observed in the samples 1 (unannealed), 4 (annealed at 214°C) and 6 (annealed at 300°C); see Table 1.

| Phase | Miller Indices | $CuK_\alpha$ | | | | $CuK_\beta$ | | | |
|---|---|---|---|---|---|---|---|---|---|
| | *h k l* | $2\theta$ tabulated[a] (deg) | $2\theta$ measured (deg) | | Intensity (counts) | $2\theta$ tabulated[a] (deg) | $2\theta$ measured (deg) | | Intensity (counts) |
| | | | Sample | Value | | | Sample | Value | |
| PtSi | 101 | 21.90 | 1 | 22.07 | weak | 19.75 | | | |
| | | | 4 | - | - | | | - | - |
| | | | 6 | - | - | | | | |
| $SiO_2$, hexagonal | | | 1 | - | - | 24.03 | 1 | - | - |
| | 101 | 26.66 | 4 | 26.56 | 365 | | 4 | 23.93 | 45 |
| | | | 6 | - | - | | 6 | - | - |
| Si | 111 | 28.44 | 1 | 28.43 | 2730 | 25.63 | 1 | 25.58 | 570 |
| | | ($K_{\alpha1}$) | 4 | 28.45 | 870 | | 4 | 25.61 | 120 |
| | | | 6 | 28.45 | 1040 | | 6 | 25.59 | 200 |
| Si | 002 | 32.96 | 1 | 32.96 | 21600 | 29.71 | 1 | 29.71 | 5100 |
| | | ($K_{\alpha1}$) | 4 | 32.95 | 15300 | | 4 | 29.71 | 3440 |
| | | | 6 | 32.96 | 20400 | | 6 | 29.71 | 4900 |
| Si | 002 | 33.04 | 1 | 33.04 | 10460 | | | | |
| | | ($K_{\alpha2}$) | 4 | 33.04 | 7120 | - | | - | - |
| | | | 6 | 33.05 | 9800 | | | | |
| Instrument | | | 1 | 33.64 | 520 | | | | |
| | - | - | 4 | 33.64 | 500 | - | | - | - |
| | | | 6 | 33.63 | 640 | | | | |
| $Pt_3Si$, monoclinic | 222 | 39.81 | 1 | 39.62 | 610 | 35.81 | 1 | 35.57 | 160 |
| Pt, cubic | 100 | 39.80 | 4 | 39.53 | 83 | | 4 | - | - |
| | | | 6 | 39.65 | 80 | | 6 | - | - |
| Si | 220 | 47.34 | 1 | 47.47 | 500 | 42.51 | 1 | 42.54 | weak |
| | | | 4 | 47.37 | 350 | | 4 | 42.43 | 60 |
| | | | 6 | 47.32 | 2400 | | 6 | 42.5 | 480 |
| Instrument | | | 1 | 54.54 | 1520 | | | | |
| | - | - | 4 | 54.55 | 1100 | - | | - | - |
| | | | 6 | 54.53 | 1420 | | | | |
| Instrument | | | 1 | 55.4 | 580 | | | | |
| | - | - | 4 | 55.40 | 470 | - | | - | - |
| | | | 6 | 55.40 | 600 | | | | |
| Si | 311 | 56.17 | 1 | 56.33 | 5910 | 50.31 | 1 | 50.30 | 150 |
| | | | 4 | 56.31 | 5100 | | 4 | 50.31 | 80 |
| | | | 6 | 56.32 | 6600 | | 6 | 50.30 | 100 |
| Si | 422 | 88.12 | 1 | 88.08 | weak | 77.79 | 1 | 77.77 | weak |
| | | | 4 | - | - | | 4 | - | - |
| | | | 6 | - | - | | 6 | - | - |
| Instrument | | | 1 | 90.93 | 230 | | | | |
| | - | - | 4 | 90.93 | 240 | - | | - | - |
| | | | 6 | 90.93 | 220 | | | | |
| Si | 333 | 94.95 | 1 | 94.95 | 60 | 83.52 | | | |
| | | | 4 | - | - | | | - | - |
| | | | 6 | - | - | | | | |
| Si | 006 | 116.64 | 1 | 116.64 | 3070 | 100.54 | 1 | 100.55 | 1340/ |
| | | ($K_{\alpha1}$) | 4 | 116.64 | 2500 | | 4 | - | - |
| | | | 6 | - | - | | 6 | - | - |
| Si | 006 | 117.10 | 1 | 117.10 | 1190 | | | | |
| | | ($K_{\alpha2}$) | 4 | 117.10 | 1120 | - | | - | - |
| | | | 6 | - | - | | | | |

[a]PDF-2 and ICSD [19–26]

Table 3. XRD reflexes observed in the samples 7, 9 and 10 (Table 1), processed at 320, 400°C and 480°C, respectively; the range of diffraction angles $2\theta$ is 15 to 60º and 71 to 115º.

| Phase | Miller Indices | CuKα | | | | CuKβ | | | |
|---|---|---|---|---|---|---|---|---|---|
| | *h k l* | $2\theta$ tabulated[a] (deg) | $2\theta$ measured (deg) Sample | Value | Intensity (counts) | $2\theta$ tabulated[a] (deg) | $2\theta$ measured (deg) Sample | Value | Intensity (counts) |
| PtSi, orthorhombic | 101 | 21.90 | 7 | 21.92 | 4150 | 19.75 | 7 | 19.75 | 630 |
| | | | 9 | 21.92 | 5700 | | 9 | 19.76 | 940 |
| | | | 10 | 21.88 | 1350 | | 10 | 19.78 | 210 |
| Si | 111 | 28.44 | 7 | 28.5[b] | 1000 | 25.63 | 7 | 25.6 | 80 |
| | | | 9 | 28.5[b] | 1000 | | 9 | 25.7 | 100 |
| | | | 10 | 28.5[b] | 950 | | 10 | 25.7 | 80 |
| PtSi, orthorhombic | 011 | 29.097 | 7 | 29.10 | 1360 | 26.22 | 7 | 26.25 | 400 |
| | | | 9 | 29.14 | 4200 | | 9 | 26.27 | 965 |
| | | | 10 | 29.15 | 7500 | | 10 | 26.27 | 1420 |
| PtSi, orthorhombic | 200 | 32.15 | 7 | 32.07 | 27800 | 28.96 | 7 | 28.91 | 3280 |
| $Pt_3Si$, monoclinic | 202 | 32.21 | 9 | 32.08 | 36340 | | 9 | 29.89 | 4000 |
| | | | 10 | 32.08 | 34700 | | 10 | 28.90 | 4200 |
| Si | 002 | 32.96 | 7 | 32.96 | 20000 | 29.71 | 7 | 29.72 | 4300 |
| | | ($K_{\alpha1}$) | 9 | 32.96 | 25000 | | 9 | 29.71 | 4200 |
| | | | 10 | 32.96 | 19000 | | 10 | 29.71 | 5500 |
| Si | 002 | 33.04 | 7 | 33.05 | 8200 | | | | |
| | | ($K_{\alpha2}$) | 9 | 33.04 | 81000 | - | - | | - |
| | | | 10 | 33.05 | 10000 | | | | |
| PtSi, orthorhombic | 111 | 33.34 | 7 | 33.38 | 200 | 30.02 | weak | | |
| $Pt_3Si$, monoclinic | 202 | 33.29 | 9 | 33.33 | 300 | | | | - |
| | | | 10 | 33.34 | 200 | | | | |
| Instrument | | | 7 | 33.83 | 390 | | | | |
| | - | - | 9 | 33.82 | 430 | - | - | | - |
| | | | 10 | 33.82 | 420 | | | | |
| PtSi, orthorhombic | 201 | 35.63 | 7 | 35.56 | 600 | 32.07 | merges with 32.08 | | |
| | | | 9 | 35.56 | 570 | | | | - |
| | | | 10 | 35.56 | 580 | | | | |
| PtSi, orthorhombic | 112 | 42.78 | 7 | 42.81 | 300 | 38.45 | weak | | |
| | | | 9 | 42.80 | 530 | | | | - |
| | | | 10 | 42.8 | 600 | | | | |
| PtSi, orthorhombic | 211 | 43.91 | 7 | 43.84 | 13200 | 39.46 | 7 | 39.41 | 2300 |
| | | | 9 | 43.88 | 21570 | | 9 | 39.45 | 4060 |
| | | | 10 | 43.90 | 23700 | | 10 | 39.44 | 4700 |
| PtSi, orthorhombic | 202 | 44.66 | 7 | 44.60 | 1840 | 40.13 | 7 | 40.09 | 340 |
| | | | 9 | 44.61 | 2690 | | 9 | 40.10 | 460 |
| | | | 10 | 44.60 | 2630 | | 10 | 40.12 | 520 |
| Si | 220 | 47.34 | 7 | 47.35 | 1800 | 42.51 | 7 | 42.54 | 300 |
| | | | 9 | 47.34 | 1500 | | 9 | 42.55 | 440 |
| | | | 10 | 47.35 | 2380 | | 10 | 42.50 | 300 |
| PtSi, orthorhombic | 020 | 50.91 | 7 | 50.94 | 2150 | 45.67 | 7 | 45.72 | 460 |
| | | | 9 | 51.01 | 2330 | | 9 | 45.73 | 1100 |
| | | | 10 | 50.98 | 8000 | | 10 | 45.72 | 1650 |
| PtSi, orthorhombic | 301 | 51.61 | 7 | 51.43 | 1570 | 46.13 | 7 | 46.16 | 445 |
| | 212 | 51.76 | 9 | 51.51 | 1140 | | 9 | 46.18 | 480 |
| $Pt_3Si$, monoclinic | 331 | 51.27 | 10 | 51.47 | 2000 | | 10 | 46.21 | 500 |
| | 310 | 51.83 | | | | | | | |
| Instrument | | | 7 | 54.54 | 1000 | | | | |
| | - | - | 9 | 54.54 | 380 | - | - | | - |
| | | | 10 | 54.54 | 1500 | | | | |
| Instrument | | | 7 | 55.32 | 360 | | | | |
| | - | - | 9 | 54.46 | 160 | - | - | | - |
| | | | 10 | 54.38 | 550 | | | | |
| Si | 311 | 56.17 | 7 | 56.31 | 4100 | 50.31 | 7 | - | - |
| | | | 9 | 56.36 | 1480 | | 9 | - | - |
| | | | 10 | 56.31 | 5840 | | 10 | 50.24 | 434 |
| PtSi, orthorhombic | 004 | 63.27 | | | | 56.53 | 7 | 56.29 | 4000 |
| | 122 | 62.82 | - | | - | 56.14 | 9 | 56.32 | 1430 |
| | | | | | | | 10 | 56.30 | 6000 |
| PtSi, orthorhombic | 411 | 74.91 | 7 | 74.75 | 1400 | 66.61 | | | |
| | | | 9 | 74.78 | 1660 | | - | | - |
| | | | 10 | 74.82 | 2000 | | | | |
| PtSi, orthorhombic | 321 | 75.45 | 7 | 75.34 | 570 | 67.07 | | | |
| | | | 9 | 75.33 | 1100 | | - | | - |
| | | | 10 | 75.34 | 630 | | | | |
| $Pt_3Si$, cubic | 311 | 82.26 | 7 | 82.34 | 120 | 72.86 | 7 | - | - |
| PtSi, orthorhombic | 031 | 82.25 | 9 | 82.35 | 90 | | 9 | - | - |
| | | | 10 | 82.44 | 350 | | 10 | 73.24 | 100 |
| PtSi, orthorhombic | 314 | 88.78 | 7 | 88.92 | 300 | 78.34 | 7 | 78.40 | 110 |
| | 420 | 89.03 | 9 | 88.94 | 470 | 78.55 | 9 | 78.51 | 300 |
| | | | 10 | 88.96 | 500 | | 10 | 78.46 | 170 |
| Instrument | | | 7 | 91.02 | 1639 | | | | |
| | - | - | 9 | 90.95 | 2780 | - | | - | - |
| | | | 10 | 90.90 | 3501 | | | | |
| Instrument | | | 7 | 91.24 | 1436 | | | | |
| | - | - | 9 | 91.25 | 1968 | - | | - | - |
| | | | 10 | 91.24 | 2649 | | | | |
| PtSi, orthorhombic | 242 | 96.60 | 7 | 96.79 | 130 | 84.78 | 7 | - | - |
| | | | 9 | 96.71 | 190 | | 9 | - | - |
| | | | 10 | 96.79 | 230 | | 10 | 84.97 | 50 |
| PtSi, orthorhombic | 512 | 104.76 | 7 | 104.92 | 110 | 91.32 | 7 | 91.34 | 40 |
| PtSi, orthorhombic | 610 | 105.01 | 9 | - | - | | 9 | 91.36 | 30 |
| | | | 10 | 104.92 | 130 | | 10 | 91.32 | 30 |
| PtSi, orthorhombic | 620 | 111.91 | 7 | 111.86 | 110 | 96.86 | merges with 96.79 | | - |
| | | | 9 | 111.91 | 50 | | | | |
| | | | 10 | 111.94 | 130 | | | | |
| Si | 006 | 116.64 | 7 | - | - | 100.54 | 7 | 100.57 | 1070 |
| | | ($K_{\alpha1}$) | 9 | 116.60 | 2880 | | 9 | 100.55 | 1270 |
| | | | 10 | 116.62 | 3200 | | 10 | 100.57 | 1880 |
| Si | 006 | 117.10 | 7 | - | - | | | | |
| | | ($K_{\alpha2}$) | 9 | 117.08 | 1360 | - | | - | - |
| | | | 10 | 117.08 | 1390 | | | | |
| PtSi, orthorhombic | 143 | 118.32 | 7 | 118.57 | 80 | 101.66 | 7 | 101.68 | 30 |
| | 004 | | 9 | 118.60 | 140 | | 9 | 101.68 | 30 |
| | | | 10 | 118.68 | 200 | | 10 | 101.92 | 10 |
| PtSi, orthorhombic | 161 | 119.98 | 7 | 120.02 | 220 | 102.86 | 7 | 102.86 | 40 |
| | | | 9 | - | - | | 9 | - | - |
| | | | 10 | 120.19 | 250 | | 10 | 102.95 | 45 |
| PtSi, orthorhombic | 451 | 126.05 | 7 | 126.19 | 70 | 107.15 | 7 | 107.20 | 30 |
| | | | 9 | - | - | | 9 | - | - |
| | | | 10 | 126.19 | 100 | | 10 | 107.20 | 50 |

[a]PDF-2 and ICSD [19–26].

[b]Poorly resolved.